\documentclass[12pt]{article}
\usepackage{a4wide}
\usepackage{amssymb}
\usepackage{graphicx}
\usepackage{comment}
\begin{document}
{\renewcommand{\thefootnote}{\fnsymbol{footnote}}
\begin{center}
{\LARGE A new type of large-scale signature change\\[2mm] in emergent modified gravity}\\
\vspace{1.5em}
Martin Bojowald,\footnote{e-mail address: {\tt bojowald@psu.edu}}
Erick I.\ Duque\footnote{e-mail address: {\tt eqd5272@psu.edu}}
and Dennis Hartmann\footnote{e-mail address: {\tt dennis.hartmann@psu.edu}}
\\
\vspace{0.5em}
Institute for Gravitation and the Cosmos,\\
The Pennsylvania State
University,\\
104 Davey Lab, University Park, PA 16802, USA\\
\vspace{1.5em}
\end{center}
}

\setcounter{footnote}{0}

\begin{abstract}
  Emergent modified gravity presents a new class of gravitational theories in
  which the structure of space-time with Riemannian geometry of a certain
  signature is not presupposed. Relying on crucial features of a canonical
  formulation, the geometry of space-time is instead derived from the
  underlying dynamical equations for phase-space degrees of freedom together
  with a crucial covariance condition. Here, a large class of spherically
  symmetric models is solved analytically for Schwarzschild-type black-hole
  configurations with generic modification functions, using a variety of
  slicings that explicitly demonstrates general covariance. For some choices
  of the modification functions, a new type of signature change is found and
  evaluated. In contrast to previous versions discussed for instance in models
  of loop quantum gravity, signature change happens on timelike hypersurfaces
  in the exterior region of a black hole where it is not covered by a
  horizon. A large region between the horizon and the signature-change
  hypersurface may nevertheless be nearly classical, such that the presence of
  a signature-change boundary around Lorentzian space-time, or a Euclidean
  wall around the universe, is consistent with observations provided signature
  change happens sufficiently far from the black hole.
\end{abstract}

\section{Introduction}

One of the motivations of modified or quantum gravity is that a consistent
theory that goes beyond general relativity at large curvature may be able to
solve important problems of the classical theory, such as the presence of
singularities. Depending on the approach, there are various expectations as to
what quantum gravity might entail, such as fundamental discreteness of space
and time that could undo the usual continuum picture of a space-time manifold
equipped with Riemannian geometry. Because a fully discrete description of
space-time is rather intractable, it is preferable to proceed more carefully
and see how the classical continuum theory could be modified by quantum or
other corrections as the curvature scale is increased. Such a treatment has
the advantage that Riemannian geometry (and therefore an unambiguous meaning
of the curvature scale) remains applicable at least for some time on the
approach to large curvature. It also makes it possible to retain a
well-defined meaning of black holes through the usual definition of a horizon.

In this way, one is led to effective line elements that, on one hand, may
include quantum or other modifications and, on the other hand, make it
possible to apply the usual concepts of curvature and black holes. However,
the application of line elements also means that covariance must be maintained
strictly at this level, even if one expects that discrete or other fundamental
space-time structures may ultimately break continuous symmetries. A line
element, classical or effective, is well-defined only if its metric
coefficients transform in such a way that any change of coordinate
differentials is compensated for exactly, implying an invariant line
element. This property cannot be tested if one works in a preferred gauge or
slicing of space-time, as often done in such models.

Within spherically symmetric models, a complete set of covariant theories with
second-order field equations has recently been derived as emergent modified
gravity \cite{HigherCov,SphSymmMinCoup}, building on previous explicit
models in which covariance could be demonstrated
\cite{SphSymmEff,SphSymmEff2}. Even in vacuum, these theories are more general
than just the classical theory even though they have the same derivative order
of equations of motion. The usual restrictions on invariant actions based on
curvature invariants can be circumvented by exploiting subtle features of the
canonical formulation of space-time theories, which turn out to be more
general than action principles because they do not require assumptions about
the space-time volume element. In particular, within emergent modified gravity
it is possible to have a space-time geometry determined by a function of the
fundamental fields that follows from an evaluation of the theory (and is in
this sense emergent), in contrast to action principles that require a
fundamental metric or tretrad field.

The general set of models is also more general than the specific examples
analyzed in \cite{SphSymmEff,SphSymmEff2}, in particular because some of the
parameter choices make it possible to have a covariant form of signature
change. The rest of this paper presents a detailed analysis of such models and
implications for black holes. Remarkably, we will find closed-form analytical
solutions even in the presence of two generic modification functions. We
therefore derive geometrical properties that are universally valid within a
large class of emergent modified gravity. The underlying theory and equations
are presented in Section~\ref{s:EMG}, followed by detailed derivations of
solutions of Schwarzschild and Painlev\'e--Gullstrand type in
Section~\ref{s:Solutions}, and a discussion of causal structure focussing on
the signature-change hypersurface in Section~\ref{s:Causal}. Unlike in
previous examples in models of loop quantum gravity \cite{Action,SigChange},
signature change in the present examples happens on a timelike hypersurface
located at low curvature. It is therefore important that our solutions are
valid for a sufficiently large class of modification function that may imply
this new type of signature change and, at the same time, are consistent with
observations in a large part of space-time outside of the black-hole
horizon. We will see that this is indeed possible.

\section{Emergent modified gravity of spherically symmetric models}
\label{s:EMG}

A generic spherically symmetric line element can be written as
\begin{equation} \label{ds}
    {\rm d}s^2= -N(t,x)^2{\rm d}t^2+ \frac{e_2(t,x)^2}{e_1(t,x)} ({\rm
      d}x+M(t,x){\rm d}t)^2+ e_1(t,x) {\rm d}\Omega^2
\end{equation}
with the lapse function $N$, the shift vector $M$, and a spatial metric
derived from components $e_1$ and $e_2$ of a densitized triad. (Without loss
of generality, we assume $e_1>0$, fixing the orientation of the spatial
triad.)  On a phase space given by the fields $(e_1,e_2)$ and canonically
conjugate momenta $(k_1,k_2)$, the dynamics is governed completely by the
diffeomorphism constraint
\begin{equation}
  D[M] = \int {\rm d} x\ M \left( k_2' e_2 - k_1 e_1' \right)
\end{equation}
and the Hamiltonian constraint  \cite{SphSymm,SPhSymmHam}
\begin{equation}
    H[N] = \int{\rm d}xN \Bigg( \frac{(e_1')^2}{8
                        \sqrt{e_1} e_2} 
    - \frac{\sqrt{e_1}}{2 e_2^2} e_1' e_2'
    + \frac{\sqrt{e_1}}{2 e_2} e_1''
    - \frac{e_2}{2 \sqrt{e_1}}
    - \frac{e_2 k_2^2}{2 \sqrt{e_1}}
    - 2 \sqrt{e_1} k_1k_2 \Bigg) \,.
\end{equation}
Classically, Hamilton's equations generated by $H[N]+D[M]$ for $e_1$ and $e_2$
show that $k_1$ and $k_2$ are related to components of extrinsic curvature of
a constant-$t$ slice in a space-time with line element (\ref{ds}).

\subsection{Covariance conditions}

At the same time, the constraints generate gauge transformations via
Hamilton's equations of $H[\epsilon^0]+D[\epsilon]$ with gauge functions
$\epsilon^0$ and $\epsilon$, whose geometrical role as hypersurface
deformations in spherically symmetric space-time is determined by the Poisson brackets
\begin{eqnarray}
  \{D[\vec{M}_1],D[\vec{M}_2]\}&=& D[{\cal L}_{\vec{M}_1}\vec{M}_2]\label{S}\\
{} \{H[N],D[\vec{M}]\} &=& -H[{\cal L}_{\vec{M}}N]\\
{} \{H[N_1],H[N_2]\} &=& D[e_1e_2^{-2}(N_1 N_2'-N_2 N_1')]\,. \label{TT}
\end{eqnarray}
These gauge transformations make sure that the line element (\ref{ds})
describes a well-defined space-time geometry irrespective of the time
coordinate $t$ chosen to define constant-$t$ hypersurfaces: When the
constraints $D[M]=0=H[N]$ and the equations of motion they generate are
satisfied, gauge transformations of the canonical theory are equivalent to
coordinate transformations of $(t,x)$ on spherically symmetric space-time
\cite{DiracHamGR,Regained}.  This classic result makes it possible to
interpret solutions of the canonical theory as space-time geometries.

Such an interpretation relies on several properties of the classical canonical
theory that may easily be broken if the constraints are modified, for instance
by possible quantum corrections. There are three broad conditions of relevant
structures being preserved: (i) The modified constraints must remain first
class, such that their mutual Poisson brackets still vanish on the constraint
surface. If this is the case, the modification does not introduce gauge
anomalies. (ii) For gauge symmetries of a modified theory to correspond to
hypersurface deformations in some space-time, the specific form of the
brackets (\ref{S})--(\ref{TT}) must be preserved. This condition is stronger
than just requiring first-class constraints because it prohibits modifications
that could, for instance, add a Hamiltonian constraint to the right-hand side
of (\ref{TT}). Such a modification would be first class, but it would not have
the correct form required for hypersurface deformations. One modification of
the brackets is nevertheless possible: The classical inverse radial metric
$q^{xx}=e_1e_2^{-2}$ in (\ref{TT}) could be replaced by a different
phase-space function, $q^{xx}_{\rm em}$. The brackets would then be compatible
with hypersurface deformations in a modified (or emergent) space-time in which
the inverse of the new function $q^{\rm em}_{xx}=1/q^{xx}_{\rm em}$ (assuming,
for now, that it is positive) provides the radial metric component. A
candidate space-time line element is then given by
\begin{equation} \label{dsem}
    {\rm d}s^2= -N^2{\rm d}t^2+ q_{xx}^{\rm em} ({\rm
      d}x+M{\rm d}t)^2+ e_1 {\rm d}\Omega^2\,.
\end{equation}
However, the condition on constraint brackets does not guarantee that the
phase-space function $q_{xx}^{\rm em}$ is subject to gauge transformations
compatible with  coordinate changes in an emergent space-time with line
element (\ref{dsem}). There is therefore a third covariance condition, (iii),
that makes sure that gauge transformations of $q_{xx}^{\rm em}$ are equivalent
to coordinate changes of a radial metric component if the constraint
equations and equations of motion are satisfied.

The three conditions are strong, but it turns out that they leave room for
modifications of the classical theory, even in vacuum without introducing
extra fields or higher derivatives. (There are also compatible matter
couplings to perfect fluids \cite{EmergentFluid} and scalar fields
\cite{EmergentScalar}.) As a new feature, they make it possible to describe
signature change in a covariant manner within a singe theory: If the classical
$e_1e_2^{-2}$ in (\ref{TT}) is replaced by a phase-space function $\gamma$
that is not positive definite, it can define a spatial metric only as
$q_{xx}^{em}=|\gamma|^{-1}$, while the compatibility of modified gauge
transformations with coordinate changes then requires the sign of $\gamma$ to
multiply $N^2$ in the time component of the metric \cite{EffLine}. In general,
the emergent space-time line element therefore reads
\begin{equation} \label{dsembeta}
    {\rm d}s^2= -{\rm sgn}(\gamma) N^2{\rm d}t^2+ \frac{1}{|\gamma|} ({\rm
      d}x+M{\rm d}t)^2+ e_1 {\rm d}\Omega^2
\end{equation}
if (\ref{TT}) is modified to
\begin{equation}
  \{H[N_1],H[N_2]\} = D[\gamma (N_1 N_2'-N_2 N_1')]\,.
\end{equation}

An interesting (though not completely general) class of modified theories can
be obtained by replacing $H[N]$ of the classical theory with a linear
combination $H[\alpha N]+D[\beta N]$ where $\alpha$ and $\beta$ are suitable
phase-space functions. The diffeomorphism constraint is left unmodified such
that the spatial structure remains classical. The fact that a linear
combination of algebra generators can modify the resulting dynamics is
somewhat counter-intuitive but, as explained in detail in
\cite{HigherCov}, it is possible because the Hamiltonian constraint
$H[N]$, by definition, generates hypersurface deformations in the normal
direction. Redefining the Hamiltonian constraint therefore changes the normal
direction $n^{\mu}$, and the latter together with the inverse spatial metric
$q^{\mu\nu}$ determines the inverse space-time metric
$g^{\mu\nu}=q^{\mu\nu}-n^{\mu}n^{\nu}$. A redefined Hamiltonian constraint may
then change the compatible space-time geometry of solutions, even though
it does not modify the constraint surface on which $H[N]=0$ and
$D[M]=0$. However, there is a well-defined space-time geometry only if our
conditions (i)--(ii) formulated above are satisfied.

These conditions, specialized to linear combinations of the Hamiltonian and
diffeomorphism constraints, have been evaluated in \cite{HigherCov}. Condition
(i) is automatically satisfied in this case. Condition (ii), requiring the
specific form of hypersurface-deformation brackets, implies that $\alpha$ and
$\beta$ are related by
\begin{equation}
  \beta(e_1,k_2)=-\frac{\sqrt{e_1}e_1'}{2e_2^2} \frac{\partial\alpha}{\partial
    k_2}- 2\sqrt{e_1}k_2 \frac{\partial\alpha}{\partial e_1'}\,.
\end{equation}
Condition (iii), imposing covariance in the sense that the resulting modified
structure function $\gamma$ transforms like an inverse radial metric, then
requires that 
\begin{equation} \label{alpha}
    \alpha(e_1,k_2) = \mu(e_1)\sqrt{1-s\lambda(e_1)^2k_2^2}
\end{equation}
with two free functions $\mu$ and $\lambda$, depending only on
$e_1$. Moreover, for later convenience, a sign factor $s=\pm 1$ has been
extracted explicitly in this equation.

Using all the conditions, the equation resulting from (ii)
leads to
\begin{equation} \label{beta}
  \beta(e_1,k_2) = \mu(e_1)\frac{\sqrt{e_1}}{2e_2^2} \frac{\partial
    e_1}{\partial x} \frac{s\lambda(e_1)^2
    k_2}{\sqrt{1-s\lambda(e_1)^2k_2^2}}
\end{equation}  
while the modified structure function equals
\begin{equation} \label{gamma}
  \gamma = \mu(e_1)^2
  \left(1+ \frac{1}{4e_2^2} \frac{s\lambda(e_1)^2}{1-s\lambda(e_1)^2k_2^2}
    \left(\frac{\partial e_1}{\partial x}\right)^2\right)
  \frac{e_1}{e_2^2}\,.
\end{equation}
We obtain the emergent radial metric
\begin{equation} \label{qeff}
  q_{xx}^{\rm em} = \mu(e_1)^{-2}
  \left|1+ \frac{1}{4e_2^2} \frac{s\lambda(e_1)^2}{1-s\lambda(e_1)^2k_2^2}
    \left(\frac{\partial e_1}{\partial x}\right)^2\right|^{-1}
  \frac{e_2^2}{e_1}
\end{equation}
and the signature factor
\begin{equation}
  \epsilon={\rm sgn}(\gamma)={\rm sgn}
  \left(1+ \frac{1}{4e_2^2} \frac{s\lambda(e_1)^2}{1-s\lambda(e_1)^2k_2^2}
    \left(\frac{\partial e_1}{\partial x}\right)^2\right)\,.
\end{equation}
These two expressions define
the emergent space-time line element
\begin{equation} \label{dsemx}
  {\rm d}s^2_{\rm em}=-\epsilon N^2{\rm d}t^2+q_{xx}^{\rm em} ({\rm
    d}x+M{\rm d}t)({\rm d}x+M{\rm d}t)+ e_1{\rm d}\Omega^2\,.
\end{equation}
The inverse space-time metric equals
\begin{equation}
    g^{\mu \nu}_{\rm em} =
    \frac{1}{q_{xx}^{\rm em}} s^\mu_x s^\nu_x
    + \frac{1}{e_1} \left( s^\mu_\vartheta s^\nu_\vartheta + \csc (\vartheta) s^\mu_\varphi s^\nu_\varphi \right)
    - n^\mu n^\nu
    \ ,
    \label{eq:Inverse metric - spherical symmetry}
\end{equation}
where 
\begin{equation}
    n^\mu = \frac{1}{N}(t^\mu - N^x s^\mu_x)
   \label{eq:Normal vector - spherical symmetry}
\end{equation}
with a spatial basis $(s^{\mu}_x,s^\mu_\vartheta,s^\mu_\varphi)$.

\subsection{The signature-change hypersurface}
\label{sec:The signature change hypersurface}

Models with $s=-1$ have not been studied in detail yet. While $s=1$ implies a
positive definite structure function, which then directly determines the
inverse spatial metric, $s=-1$ may allow for ranges of $x$ in which $\gamma$
is negative. The emergent space-time then has Euclidean signature in such a
region, separating it from Lorentzian signature at positive $\gamma$ by a
signature-change hypersurface in space-time. Such hypersurfaces are defined by
$\gamma(t_{\rm sc},x_{\rm sc})=0$, which may have disjoint solutions for
$(t_{\rm sc},x_{\rm sc})$, implying multiple signature-change hypersurfaces in
general. A signature-change hypersurface may be timelike or spacelike
depending on where it appears relative to horizons. In black-hole models, a
signature-change hypersurface that occurs in a static (exterior) region is
timelike because $\gamma$ can depend only on the spatial coordinate $x$, which
lies in a discrete set of values determined by $\gamma(x_{\rm sc})=0$. In a
spatially homogeneous model for a black-hole interior, $\gamma$ depends only
on the time coordinate $t$, such that a signature-change hypersurface in this
region is spacelike, determined by $t=t_{\rm sc}$ with a solution $t_{\rm sc}$
of $\gamma(t_{\rm sc})=0$. We will see that $s=-1$ in our models can only lead
to timelike signature-change hypersurfaces at a unique value of $x_{\rm
  sc}$. The following discussion of general properties is based on this
outcome, but similar statements can easily be made for spacelike
signature-change hypersurfaces as well.

In a static region, the occurrence of a timelike signature-change hypersurface
requires that the structure function vanishes at a certain value of
$x=x_{\rm sc}$. (If the region is not static but the signature-change
hypersurface remains timelike, it is always possible to introduce local
coordinates such that the hypersurface is defined by a constant value of the
radial coordinate.) Starting in a range of $x$-values for which the emergent
space-time metric has Lorentzian signature, inspection of the inverse metric
(\ref{eq:Inverse metric - spherical symmetry}) reveals that near the
signature-change hypersurface, space-time degenerates into a family of
$(2+1)$-dimensional geometries with inverse metric
\begin{equation}
    g^{\mu \nu}_{\rm em} \approx
    \frac{1}{x_{\rm sc}^2} \left( s^\mu_\vartheta s^\nu_\vartheta + \csc^2
      (\vartheta) s^\mu_\varphi s^\nu_\varphi \right) 
    - n^\mu n^\nu
    \label{eq:Inverse metric - spherical symmetry - near signature change}
\end{equation}
and topology $\mathbb{R} \times S^2$. Approaching the hypersurface from the
Euclidean region, the signature-change hypersurface is the limiting case
of a family of spacelike hypersurfaces.

It then follows that the radial component of the emergent metric (\ref{qeff})
diverges at $x_{\rm sc}$.  This conclusion holds irrespective of the gauge or
coordinate system used because $\det(g^{-1})=0$ is a coordinate invariant
statement. Therefore, there is no coordinate choice that can remove this
divergence of a metric component, implying a physical singularity according to the
standard definition.  However, invariant objects such as the Ricci scalar are
not necessarily singular at a signature-change hypersurface.  Our model in
spherical symmetry derived below provides an example of a signature-change
hypersurface with a non-singular geometry.

The standard definition of geodesic incompleteness might also suggest a
physical singularity because timelike geodesics from the Lorentzian region
cannot be extended as timelike geodesics into the Euclidean region. However,
there may be extensions to spacelike geodesics if it is possible to use the
final values of a timelike geodesic in the Lorentzian region as initial
conditions for a spacelike geodesic in the Euclidean region. An important
question related to geodesic completeness is whether such an extension is
unique, which requires a well-defined tangent vector at the hypersurface as
well as a continuous set of coordinate transformations that can be applied in
a region across the signature-change hypersurface. Details of such extensions
require specific models, but the main challenging property can be inferred
from the behavior of the space-time metric that gives rise to signature change
on a timelike hypersurface.

Assuming a timelike signature-change hypersurface at $x=x_{\rm sc}$, the
radial component $q_{xx}^{\rm em}$ of the metric diverges at this
value. Normalization of the tangent vector of a geodesic approaching
the hypersurface,
\begin{equation}
  -1= -N^2 \left(\frac{{\rm d}t}{{\rm d}\tau}\right)^2+q_{xx}^{\rm
    em}\left(\frac{{\rm d}x}{{\rm d}\tau}\right)^2+ e_1\left(\frac{{\rm
        d}\vartheta}{{\rm d}\tau}\right)^2+ e_1\sin^2\vartheta
  \left(\frac{{\rm d}\varphi}{{\rm d}\tau}\right)^2
\end{equation}
then requires that $v^x={\rm d}x/{\rm d}\tau$ approaches zero or that some of
the other velocity components diverge at the signature-change hypersurface. In
both cases, the geodesic is asymptotically tangent to the hypersurface and
does not provide a unique final direction into the Euclidean region on the
other side of the hypersurface. A similar argument follows from lightlike
geodesics, which in the radial case require a divergent
\begin{equation}
  \frac{{\rm d}t}{{\rm d}x}=\sqrt{\frac{q_{xx}^{\rm em}}{N}}
\end{equation}
at the hypersurface. In our specific models we will show that the hypersurface
may nevertheless be reached at a finite distance from an interior point of the
complete space-time manifold, including the Euclidean region.

\subsection{Hamiltonian constraints}

Using the explicit solutions for $\alpha$ and $\beta$, the new Hamiltonian
constraint is given by the expression
\begin{eqnarray}
    H^{{\rm (new)}}[N] &=& \int{\rm d}xN\mu\sqrt{1 - s \lambda^2 k_2^2} \Bigg( \left( \frac{1}{8
                        \sqrt{e_1} e_2} - s \lambda^2 \frac{\sqrt{e_1}}{2
                        e_2^2} \frac{k_1k_2}{1 - s \lambda^2
                        k_2^2} \right) (e_1')^2 
    \nonumber\\
    &&\qquad
    - \frac{\sqrt{e_1}}{2 e_2^2} e_1' e_2'
    + \frac{\sqrt{e_1}}{2 e_2} e_1''
    + s \lambda^2 \frac{\sqrt{e_1}}{2 e_2^2} \frac{e_2k_2}{1 - s \lambda^2 k_2^2} e_1' k_2'
    \nonumber\\
    &&\qquad
    - \frac{e_2}{2 \sqrt{e_1}}
    - \frac{e_2 k_2^2}{2 \sqrt{e_1}}
    - 2 \sqrt{e_1} k_1k_2 \Bigg) 
    \label{eq:Modified constraint}
\end{eqnarray}
for given  $s$, $\mu$ and $\lambda$.

The form of functions $\alpha$, $\beta$ and $\gamma$ makes use of the
phase-space variables $(e_1,e_2)$ and $(k_1,k_2)$ initially obtained in the
classical theory. However, modifications of equations of motion and of the
structure function imply that $(e_1,e_2)$ no longer are densitized-triad
components of the emergent spatial metric, and $(k_1,k_2)$ are no longer
directly related to components of extrinsic curvature. It is therefore
possible to apply canonical transformations, introducing further changes in
the phase-space dependence of $\alpha$, $\beta$ and $\gamma$. Such
transformations do not change physical or geometrical implications, but they
may sometimes be convenient for solving or interpreting equations. The
specific versions (\ref{alpha}), (\ref{beta}) and (\ref{gamma}) are unique up
to canonical transformations, provided modifications happen only by replacing
the Hamiltonian constraint with a suitable linear combination of the classical
constraints. As an example, the models analyzed in
\cite{SphSymmEff,SphSymmEff2} are equivalent to our case of $s=+1$ and
constant $\mu$ and $\lambda$ with a specific relationship between these two
constants, up to a canonical transformation of $(e_2,k_2)$. Canonical
transformations can be used to extend these models to non-constant $\lambda$,
and to describe the case of $s=-1$ by similar means.

\subsubsection{Periodic variables: Lorentzian case}

For the case of $s=1$, we perform the canonical transformation
\begin{eqnarray}
    k_2 = \frac{\sin (\lambda \tilde{k}_2)}{\lambda}
    \quad &,& \quad
    e_2 = \frac{\tilde{e}_2}{\cos (\lambda \tilde{k}_2)}
    \ , \nonumber\\
    k_1 =
    \tilde{k}_1
    + \frac{\tilde{e}_2}{\cos (\lambda \tilde{k}_2)} \frac{\partial}{\partial e_1} \left(\frac{\sin (\lambda k_2)}{\lambda}\right)
    \quad &,&\quad
    e_1 = \tilde{e}_1
\end{eqnarray}
where the new variables are written with a tilde.
The Hamiltonian constraint (\ref{eq:Modified constraint}) then becomes
\begin{eqnarray}
    H^{{\rm (c)}}_+ [N] &=& \int{\rm d}x\ N \mu \Bigg(
    \left( \frac{\cos^2(\lambda k_2)}{8
    \sqrt{e_1} e_2} - \lambda^2 \frac{\sqrt{e_1}}{2 e_2^2} \left(k_1
    + e_2 k_2 \frac{\partial \ln \lambda}{\partial e_1} \right) \frac{\sin (2 \lambda k_2)}{2 \lambda} \right) (e_1')^2 
    \nonumber\\
    &&\qquad
    - \frac{\sqrt{e_1}}{2 e_2^2} e_1' e_2' \cos^2(\lambda k_2)
    + \frac{\sqrt{e_1}}{2 e_2} e_1'' \cos^2(\lambda k_2)
    - \frac{e_2}{2 \sqrt{e_1}}
    - \frac{e_2}{2 \sqrt{e_1}} \frac{\sin^2 (\lambda k_2)}{\lambda^2}
    \nonumber\\
    &&\qquad
    - 2 \sqrt{e_1} \left( k_1 \frac{\sin (2 \lambda k_2)}{2 \lambda}
    + e_2 \left(\frac{\sin (2\lambda k_2)}{2 \lambda} k_2 - \frac{\sin^2 (\lambda k_2)}{\lambda^2}\right) \frac{\partial \ln \lambda}{\partial e_1} \right) \Bigg) 
\end{eqnarray}
where we have dropped the tilde for the sake of convenience, with structure function
\begin{eqnarray}
    q^{x x}_{{\rm (c)} +} &=& 
    \mu^2 \cos^2 (\lambda k_2) \left( 1 + \lambda^2 \left(\frac{e_1'}{2 e_2}\right)^2\right) \frac{e_1}{e_2^2}
    \,.
\end{eqnarray} 
This transformation replaces square roots by trigonometric functions, but the
dependence on $k_2$ is not periodic unless $\lambda$ does not depend on
$e_1$. A second canonical transformation,
\begin{eqnarray}
    k_2 &=&
    \frac{\bar{\lambda}}{\lambda} \tilde{k}_2
    \ ,
    \hspace{4cm}
    e_2 = \frac{\lambda}{\bar{\lambda}} \tilde{e}_2
    \ ,
    \nonumber\\
    k_1 &=&
    \tilde{k}_1
    - \tilde{e}_2 \tilde{k}_2 \frac{\partial \ln \lambda}{\partial \tilde{e}_1}
    \ ,
    \hspace{2cm}
    e_1 = \tilde{e}_1
    \label{eq:Canonical transformation to periodic variables}
\end{eqnarray}
where $\bar{\lambda}$ is an arbitrary non-zero constant, can be used to make
the Hamiltonian constraints strictly periodic in $k_2$:
\begin{eqnarray}
    H^{{\rm (cc)}}_+ [N] &=& \int{\rm d}x\ N \frac{\bar{\lambda}}{\lambda} \mu \Bigg[
    \left( \left( \frac{1}{8
    \sqrt{e_1} e_2}
    - \frac{\sqrt{e_1}}{2 e_2} \frac{\partial \ln \lambda}{\partial e_1} \right) \cos^2(\bar{\lambda} k_2)
    - \bar{\lambda}^2 \frac{\sqrt{e_1}}{2 e_2^2} k_1 \frac{\sin (2 \bar{\lambda} k_2)}{2 \bar{\lambda}} \right) (e_1')^2
    \nonumber\\
    &&\qquad
    - \frac{\sqrt{e_1}}{2 e_2^2} e_1' e_2' \cos^2(\bar{\lambda} k_2)
    + \frac{\sqrt{e_1}}{2 e_2} e_1'' \cos^2(\bar{\lambda} k_2)
    - \frac{e_2}{2 \sqrt{e_1}}
    - \frac{e_2}{2 \sqrt{e_1}} \frac{\sin^2 (\bar{\lambda} k_2)}{\bar{\lambda}^2}
    \nonumber\\
    &&\qquad
    - 2 \sqrt{e_1} \left( k_1
    - e_2 \frac{\tan (\bar{\lambda} k_2)}{\bar{\lambda}} \frac{\partial \ln \lambda}{\partial e_1} \right) \frac{\sin (2 \bar{\lambda} k_2)}{2 \bar{\lambda}} \Bigg]
    \label{eq:Modified constraint - periodic variables}
\end{eqnarray}
where we have again dropped the tilde,
with structure function
\begin{eqnarray}
    q^{x x}_{{\rm (cc)} +} &=& 
    \frac{\bar{\lambda}^2}{\lambda^2} \mu^2 \cos^2 (\bar{\lambda} k_2) \left( 1 + \bar{\lambda}^2 \left(\frac{e_1'}{2 e_2}\right)^2\right) \frac{e_1}{e_2^2}
    \,.
    \label{eq:Modified structure function - periodic variables}
\end{eqnarray}

The constraint (\ref{eq:Modified constraint - periodic variables}) and its
structure function (\ref{eq:Modified structure function - periodic variables})
are periodic in the new $k_2$. In this sense, they are related to models of
loop quantum gravity in which this periodicity is usually interpreted as a
necessary requirement for gauge theories based on holonomies. However, the
specific terms in (\ref{eq:Modified constraint - periodic variables}) are
different from most models that have been considered in this context, and in
past developments of spherically symmetric loop quantum gravity it has been
tacitly assumed that the classical $e_2^2/e_1$ still describes a meaningful
radial metric. As shown by emergent modified gravity, this assumption and any
deviations of a modified Hamiltonian constraint from (\ref{eq:Modified
  constraint - periodic variables}) violate general covariance and do not
imply reliable effective line elements.

\subsubsection{Hyperbolic variables: Signature-change case}

The constraint (\ref{eq:Modified constraint}) with $s=-1$ is mathematically
equivalent to the case $s=1$ with the substitution $\lambda \to i \lambda$.
Furthermore, in the canonical transformations used in the previous sections we
may replace trigonometric functions with hyperbolic ones.  Therefore, the case
$s=-1$ with hyperbolic canonical transformations can simply be expressed as
(\ref{eq:Modified constraint - periodic variables}) with the substitutions
$\lambda \to i \lambda$ and $\bar{\lambda} \to i \bar{\lambda}$:
\begin{eqnarray}
    H^{{\rm (cc)}}_- [N] &=& \int{\rm d}x\ N \frac{\bar{\lambda}}{\lambda} \mu \Bigg[
    \left( \left( \frac{1}{8
    \sqrt{e_1}}
    - \frac{\sqrt{e_1}}{2} \frac{\partial \ln \lambda}{\partial e_1} \right) \cosh^2(\bar{\lambda} k_2)
    + \bar{\lambda}^2 \frac{\sqrt{e_1}}{2} \frac{k_1}{e_2} \frac{\sinh (2 \bar{\lambda} k_2)}{2 \bar{\lambda}} \right) \frac{(e_1')^2}{e_2}
    \nonumber\\
    &&\qquad
    + \frac{\sqrt{e_1}}{2} \left( \frac{e_1''}{e_2} - \frac{e_1' e_2'}{e_2} \right) \cosh^2(\bar{\lambda} k_2)
    - \frac{e_2}{2 \sqrt{e_1}} \left( 1 + \frac{\sinh^2 (\bar{\lambda} k_2)}{\bar{\lambda}^2} \right)
    \nonumber\\
    &&\qquad
    - 2 \sqrt{e_1} \left( k_1
    - e_2 \frac{\tanh (\bar{\lambda} k_2)}{\bar{\lambda}} \frac{\partial \ln \lambda}{\partial e_1} \right) \frac{\sinh (2 \bar{\lambda} k_2)}{2 \bar{\lambda}} \Bigg]
    \label{eq:Modified constraint - hyperbolic variables}
\end{eqnarray}
with structure function
\begin{eqnarray}
    q^{x x}_{{\rm (cc)} -} &=& 
    \frac{\bar{\lambda}^2}{\lambda^2} \mu^2 \cosh^2 (\bar{\lambda} k_2) \left( 1 - \bar{\lambda}^2 \left(\frac{e_1'}{2 e_2}\right)^2\right) \frac{e_1}{e_2^2}
    \,.
    \label{eq:Modified structure function - hyperbolic variables}
\end{eqnarray}
The case of $s=-1$, studied as the main example in what follows, may therefore
be interpreted as a hyperbolic version of covariant models for spherically
symmetric loop quantum gravity.

\section{Solutions}
\label{s:Solutions}

Canonical equations of gravitational theories provide unique solutions only if
certain gauge choices are made that specify the coordinate frame or slicing in
which the corresponding space-time geometry is expressed. The main classical
slicing conditions used in spherically symmetric models can be
generalized to emergent modified gravity, as shown in this section.

\subsection{Schwarzschild-like exterior}

We will be using space-time solutions in various gauges when we analyze
geodesics and other properties of emergent modified gravity. For constant
$\mu$ and $\lambda$, the case of $s=1$ has been analyzed in
\cite{SphSymmMatter,SphSymmMatter2}, while the case of $s=-1$ is been
discussed briefly in \cite{CovColl}. In particular, the latter contribution shows the
possibility of signature change for $s=-1$ at large $x$ in the exterior region
of a black hole. However, this is possible only if $\lambda>1$, implying
significant modifications of gravity even in intermediate ranges of $x$ that
should be directly accessible by observations. It is therefore important to
confirm that signature change is still possible if $\lambda$ is no longer
constant and may increase from small values in observationally accessible
regimes to larger values at the outer fringes of the universe. One of the main
results of the present paper is that this is indeed possible. We will see that
most of the calculations for constant $\lambda$ go through with only minimal
changes if $\lambda$ is not constant. Our derivations in the remainder of
this section are based on the form (\ref{eq:Modified constraint}) of the
Hamiltonian constraint.

We first compute the line element of a static region of space-time suitable
for the exterior of a non-rotating black hole.  We directly obtain $M=0$,
$\dot{e}_1=0=\dot{e}_2$ and therefore $k_1=0=k_2$, which allows us to fix the
spatial gauge by declaring that $e_1(x)=x^2$. All these equations take their
classical form, and with vanishing $k$-terms for static configurations, the
Hamiltonian constraint is classical too (up to an additional multiplier of
$\mu$) and implies $e_2(x)=x/\sqrt{1-2m/x}$ where, based on the position of
the horizon, the integration constant $m$ turns out to have the same
interpretation as mass as in the classical solution.

There are additional consistency conditions from the requirement that
$\dot{k}_1=0$ and $\dot{k}_2=0$ for static behavior are compatible with the
equations of motion. They imply almost the same result
as in the classical theory,
\begin{equation}
  N\alpha\mu =\sqrt{1-\frac{2m}{x}}
\end{equation}
where $\alpha$ is a constant and can be considered a rescaling of the time
coordinate. This parameter can be used to cancel $\mu$ only if the latter is
constant, but not if it depends on $e_1$ and therefore on $x$. For
non-constant $\mu$, $m$ retains its interpretation of mass if the latter is
defined via the Schwarzschild radius at $2m$. (If $\mu$ is asymptotically
constant such that $\mu-{\rm const}$ falls off faster than $1/x$, interpreting
$m$ as the mass would also be consistent with Newton's potential. However,
depending on $s$ and $\lambda$, the asymptotic limit can be very non-classical
and may no longer be a suitable indicator of the mass.)

None of the solutions for phase-space variables are significantly
modified in this gauge. However, the emergent space-time metric, and therefore
space-time geometry, does have non-trivial corrections. For covariance of this
gauge within a well-defined space-time geometry, the emergent space-time
metric must be compatible with the full modified constraint and not just with
its static restriction in which all $k$-terms disappear. The resulting radial
metric component, given in general by (\ref{qeff}), evaluates to
\begin{equation}\label{eq: signature change structure function}
  q_{xx}^{\rm em} = \frac{\epsilon e_2^2}{\mu^2e_1(1+s\lambda^2 x^2/e_2^2)}
\end{equation}
in the present case, with $k_2=0$ and $e_1=x^2$ and the
signature factor
\begin{equation}
  \epsilon = {\rm sgn} (1+s\lambda^2x^2/e_2^2)\,.
\end{equation}
We obtain the emergent space-time line element 
\begin{eqnarray} \label{Schwarzschild}
  {\rm d}s^2_{\rm em} &=&
  -\epsilon\left(1-\frac{2m}{x}\right)\frac{{\rm d}t^2}{\alpha^2 \mu^2} + \frac{\epsilon {\rm d}x^2/\mu^2}{(1-2m/x)(1+s\lambda^{2}(1-2m/x))}+ 
  x^2{\rm d}\Omega^2
 \\
  &=&
  -\epsilon \left(1-\frac{2m}{x}\right)\frac{{\rm d}t^2}{\alpha^2\mu^2} 
  + \frac{-s \epsilon}{\alpha^2 \mu^2 (\lambda^{2}+s)}
  \left(1-\frac{2m}{x}\right)^{-1} \left(\frac{X_\lambda}{x}-1\right)^{-1} {\rm d}x^2
  + x^2{\rm d}\Omega^2 \nonumber
\end{eqnarray}
if we introduce the function
\begin{equation}\label{Xlambda}
  X_{\lambda}(x) = \frac{2 m \lambda(x)^2}{\lambda(x)^{2}+s}\,.
\end{equation}
The signature-change hypersurface is located at $x=x_{\lambda}$ with an
implicit equation
\begin{equation}\label{eq:SigChangePos}
  x_{\lambda} =X_{\lambda}(x_{\lambda})= \frac{2 m \lambda(x_{\lambda})^2}{\lambda(x_{\lambda})^{2}+s}
\end{equation}
for $x_{\lambda}$, provided $\lambda^2+s\not=0$. (For constant $\lambda$,
$X_{\lambda}$ is constant and (\ref{eq:SigChangePos}) directly defines
$x_{\lambda}$ in terms of $\lambda$.)
The signature factor evaluates to
\begin{eqnarray}
    \epsilon \equiv {\rm sgn} \left( - s (\lambda^{2}+s) \left(\frac{X_{\lambda}}{x}-1\right) \right)
    \ .
\end{eqnarray}

For $s=1$, $x_{\lambda}<2m$ is not in the static exterior and $\epsilon={\rm
  sgn}(\mu^2(1+\lambda^2)(1-X_{\lambda}/x))=1$ for $x>2m$, such that there is
no signature change in this case. The resulting line element 
\begin{equation}
   {\rm d}s^2_{\rm em} = - \left(1-\frac{2m}{x}\right)\frac{{\rm d}t^2}{\alpha^2 \mu^2} 
  + \frac{1}{\mu^2 (1+\lambda^{2})}
  \left(1-\frac{2m}{x}\right)^{-1} \left(1-\frac{X_\lambda}{x}\right)^{-1} {\rm d}x^2
  + x^2{\rm d}\Omega^2
\end{equation}
is asymptotically flat in a strict sense only if $\mu$ and $\lambda$ are
asymptotically constant for $x\gg 2m$.  (If this condition is not satisfied,
the line element is quasi-asymptotically flat \cite{QuasiAsFlat}; see also
\cite{TransCommAs}.)

If $s=-1$, the emergent spatial metric is positive definite if $\lambda<
1$. If $\lambda=1$ in a region of space,
$\epsilon={\rm sgn}(-2s\lambda^2m/x)=1$ and the line element is given by
\begin{equation}
  {\rm d}s^2_{\rm em}= -\left(1-\frac{2m}{x}\right)\frac{{\rm
      d}t^2}{\alpha^2 \mu^2} +
 \frac{x/(2m)}{1-2m/x} \frac{ {\rm  d}x^2}{\mu^2\lambda^{2}}+   
  x^2{\rm d}\Omega^2
\end{equation}
If $\lambda>1$, we have $\epsilon=1$ for $x < x_{\lambda} = 2 m \lambda^2 /
(\lambda^2-1)$, with a Loretzian-signature line element
\begin{equation}\label{SchwarzschildSig-Lorentzian region}
  {\rm d}s^2_{\rm em} = -\left(1-\frac{2m}{x}\right) \frac{{\rm d}t^2}{\alpha^2 \mu^2}+
  \frac{{\rm d}x^2/(\mu^2(\lambda^2-1))}{(1-2m/x)(X_{\lambda}/x-1)}+ x^2{\rm
    d}\Omega^2\,,
\end{equation}
and $\epsilon=-1$ 
for $x > x_{\lambda}$ with a Euclidean-signature line element
\begin{equation} \label{SchwarzschildSig}
  {\rm d}s^2_{\rm em} = \left(1-\frac{2m}{x}\right) \frac{{\rm d}t^2}{\alpha^2 \mu^2}
  + \frac{{\rm d}x^2/(\mu^2(\lambda^2-1))}{(1-2m/x)(1-X_{\lambda}/x)}+ x^2{\rm d}\Omega^2\,.
\end{equation}
There is therefore Euclidean-signature 4-dimensional space surrounding
Lorentzian space-time containing a black hole. For non-constant $\lambda$, the
Lorentzian geometry can be compatible with observations if $\lambda$ is
sufficiently small for a suitable range of $x$. If $\lambda$ grows beyond the
value one for larger $x$, we obtain the signature-change region.

The Ricci scalar of (\ref{SchwarzschildSig-Lorentzian region}) is given by
\begin{eqnarray}
    \mathcal{R} &=& \left( \frac{2}{x^2} + \left(\frac{2}{x^2} - \frac{3 m
                    x_\lambda}{x^4}\right) \mu^2 (\lambda^2 - 1) \right)  
    + \mu^2 (\lambda^2 - 1) \frac{(3 m-2 x)}{x^3} X_\lambda'
    \nonumber\\
    &&+ \frac{(\lambda^2 - 1) (x-2 m)}{2 x^2} (\mu^2)' X_\lambda'
    + \frac{(2 m-x) (x-X_\lambda)}{2 x^2} (\mu^2)' (\lambda^2 - 1)'
    \nonumber\\
    &&
    + \frac{(\lambda^2 - 1) (2 m-4 x+3 X_\lambda)}{2 x^2} (\mu^2)'
    - \frac{(\lambda^2 - 1) (2 m-x) (x-X_\lambda)}{\mu^2 x^2} ((\mu^2)')^2
    \nonumber\\
    &&
    + \frac{(\lambda^2 - 1) (2 m-x) (x-X_\lambda)}{x^2} (\mu^2)''
    - \frac{(3 m-2 x) (x-X_\lambda)}{x^3} \left(\mu^2 (\lambda^2 - 1)\right)'
\end{eqnarray}
where the primes denote $x$-derivatives.
The Ricci scalar remains finite at the signature-change hypersurface at $x=x_{\lambda}$.
The Kretschmann scalar $K\equiv R_{\mu \nu \alpha \beta} R^{\mu \nu \alpha
  \beta}$ evaluates to
\begin{eqnarray}
  &&K\nonumber\\
    &=& \frac{\mu^4 (\lambda^2 - 1)^2}{x^{8}} \Biggl( \frac{4 x^2}{\mu^4 (\lambda^2 - 1)^2}
    + \frac{8 x^2}{\mu^2 (\lambda^2 - 1)} \left(x-2 m\right) \left(x-X_\lambda\right)
    + 4 x^2 \left( x^2 - 4 m x + 12 m^2 \right)
    \nonumber\\
    &&
    - 8 x \left( x^2 - 5 m x + 15 m^2 \right) X_\lambda
    + \left( 6 x^2 - 32 m x + 81 m^2 \right) X_\lambda^2 \Biggr)
    \nonumber\\
    &&
    + \frac{4 \lambda  \left(\lambda ^2-1\right) \mu ^4 (x-X_\lambda) \left(3 m^2 (7 X_\lambda-4 x)+4 m x (x-3 X_\lambda)+2 x^2 X_\lambda\right)}{x^7} \lambda'
    \nonumber\\
    &&
    + \frac{4 \lambda ^2 \mu ^4 \left(9 m^2-8 m x+2 x^2\right) (x-X_\lambda)^2}{x^6} (\lambda')^2
    \nonumber\\
    &&
    + \frac{8 \lambda  \left(\lambda ^2-1\right) \mu ^3 (x-X_\lambda) \left(2 m^2+m (5 X_\lambda-6 x)+2 x (x-X_\lambda)\right)}{x^5} \lambda' \mu'
    \nonumber\\
    &&
    + \frac{8 \lambda ^2 \mu ^3 m (2 m-x) (x-X_\lambda)^2}{x^5} (\lambda')^2 \mu'\nonumber\\
&&    - \frac{4 \lambda  \left(\lambda ^2-1\right) \mu ^2 (2 m-x)
   (x-X_\lambda) (6 m x-8 m X_\lambda+x X_\lambda)}{x^5} \lambda' (\mu')^2 
    \nonumber\\
    &&+ \frac{2 \left(\lambda^2-1\right)^2 \mu ^3 \left(2 m^2 \left(8 x^2-14 x X_\lambda+7 X_\lambda^2\right)+m x X_\lambda (11 x_\lambda-12 x)+4 x^2 X_\lambda (x-X_\lambda)\right)}{x^7} \mu'
    \nonumber\\
    &&
    + \frac{K_1(x)}{x^6} (\mu')^2 + \frac{K_2(x)}{x^6}  \mu''
    \nonumber\\
    &&+ \frac{\mu^4(\lambda^2-1)^2 (x-2 m)^2}{4 \alpha^4 \mu^4 x^4}
       ((\mu^2(\lambda^2-1))')^2 ((X_\lambda)')^2\nonumber\\
  &&
    + \frac{\mu^4(\lambda^2-1)^2 m (2 m-x)}{\alpha^2\mu^2 x^5} (\mu^2(\lambda^2-1))' ((X_\lambda)')^2
    \nonumber\\
    &&
    + \frac{\mu^4(\lambda^2-1)^2 \left(9 m^2-8 m x+2 x^2\right)}{x^6}
       ((X_\lambda)')^2\nonumber\\
  &&
    - \frac{\mu^4(\lambda^2-1)^2 (2 m-x) (x-X_\lambda) \left(2 \alpha^2 \mu^2  m+x (2 m-x) (\alpha^2 \mu^2)'(x)\right)}{\alpha^2 \mu^4 x^5} X_\lambda' (\mu^2)''
    \nonumber\\
    &&
    + \frac{K_3(x)}{2 \mu^2 x^6} X_\lambda' (\mu^2(\lambda^2-1))'
    \nonumber\\
    &&
    + \frac{3 \mu^4(\lambda^2-1)^2 (x-2 m)^2 (x-X_\lambda)}{2 \mu^6 x^4}
       X_\lambda' ((\mu^2)')^3\nonumber\\
  &&
    + \frac{\mu^4(\lambda^2-1)^2 (2 m-x) (2 m (6 x-7 X_\lambda)+x X_\lambda)}{2 \mu^2 x^5} X_\lambda' ((\mu^2)')^2
    \nonumber\\
    &&
    + \mu^4(\lambda^2-1)^2 \frac{2 m \left(7 m x-9 m X_\lambda-2 x^2+3 x X_\lambda\right)}{\mu^2 x^6} X_\lambda' (\mu^2)'
    \nonumber\\
    &&
    + \frac{2 \mu^4(\lambda^2-1)^2 \left(3 m^2 (4 x-7 X_\lambda)-4 m x (x-3 X_\lambda)-2 x^2 X_\lambda\right)}{x^7} X_\lambda'
\end{eqnarray}
with
\begin{eqnarray}
  K_1(x) &=& \left(\lambda^2-1\right)^2 \mu ^2 \left(X_\lambda^2 \left(140
             m^2-96 m x+17 x^2\right)-4 x X_\lambda \left(62 m^2-43 m x+8
             x^2\right)\right.\nonumber\\
  &&\left.+16 x^2 \left(7 m^2-5 m x+x^2\right)\right) \nonumber\\
  K_2(x) &=&-4 \left(\lambda^2-1\right)^2 \mu^3 m (2 m-x) (4 x-5 X_\lambda)
             (x-X_\lambda)\nonumber\\
  &&
    + 8 \lambda  \left(\lambda ^2-1\right) \mu^3 m (2 m-x) (x-X_\lambda)^2
             x\lambda' \nonumber\\
  K_3(x) &=& (\lambda^2-1) (x-X_\lambda) \left(x (x-2 m) (\mu^2)' \left(4
             \mu^2  m+x (2 m-x) (\mu^2)'\right)\right.\nonumber\\
  &&\left.-4 \mu^4 \left(9 m^2-8 m x+2
             x^2\right)\right) 
             \end{eqnarray}

This expression is also finite at the signature-change hypersurface where it takes the value
\begin{eqnarray}
    K |_{x=x_\lambda} &=& \frac{1}{4 x_\lambda^6}
    \Biggl(
    8 \mu^4(\lambda^2-1)^2 (x_\lambda-2 m)^2 \left(X_\lambda'(x_{\lambda})-1\right)^2+16 x_\lambda^2
    \nonumber\\
    &&
    \frac{\mu^4(\lambda^2-1)^2 \left(X_\lambda'(x_{\lambda})-1\right)^2
       \left(x_\lambda (x_\lambda-2 m) (\mu^2)'-2 \mu^2 m\right)^2}{\mu^4} 
    \Biggr)
    \,.
\end{eqnarray}
Both scalars
retain their classical divergence at $x=0$. (In the next subsection we will
see that the exterior solution can be extended to $x<2m$ in the usual way by
flipping the role of radial and time coordinates.)

We conclude that the singularity in the metric
(\ref{SchwarzschildSig-Lorentzian region}) at $x=x_{\lambda}$ may be
consistent with an interpretation as a coordinate singularity.

\subsection{Homogeneous interior}

In the classical Schwarzschild solution, the interior geometry for $x<2m$ can
be obtained from the exterior solution by flipping the roles of $t$ and $x$ as
time and space, respectively. In emergent modified gravity, only the spatial
part of the metric contains additional terms based on the covariance
condition, while the time component, through the lapse function, may be
modified only indirectly based on the equations it has to solve in a given
gauge. It is therefore not clear that simply flipping $t$ and $x$ correctly
transfers additional terms in the emergent spatial metric from the radial part
to the time component. In the present case, an explicit independent derivation of the
interior solution demonstrates that the classical procedure nevertheless applies.

As part of the gauge choice for a Schwarzschild-type interior, we assume that
all fields, $e_1$, $e_2$, $k_1$, $k_2$ and $N$, depend only on a time
coordinate $T$, and that $M = 0$. The function $e_1$ was fixed by a simple
gauge choice in the exterior. Flipping the coordinates is possible only if we
have the same simple choice in the interior, but now as a dependence 
$e_1(T) = T^2$ on the new time coordinate $T$. The modification functions
$\mu$ and $\lambda$ may therefore be time dependent.

This $e_1$ has to be compatible with the equation of motion
\begin{equation}
    \dot{e}_1 = - 2 \mu N \sqrt{e_1} k_2 \sqrt{1 - s \lambda^2 k_2^2}
    \ ,
\end{equation}
which, using $\dot{e}_1=2T=2\sqrt{e_1}$, relates $N$ to $k_2$ by
\begin{equation}
    N = - \frac{1}{\mu k_2 \sqrt{1 - s \lambda^2 k_2^2}}
    \,.
\end{equation}
The equation of motion for $k_2$,
\begin{equation}
    \dot{k}_2 = \frac{\mu N}{2 \sqrt{e_1}} \sqrt{1-s \lambda^2 k_2^2} \left(1+k_2^2\right) = - \frac{1}{2 T} \frac{1+k_2^2}{k_2}
    \ ,
\end{equation}
can then be solved directly by 
\begin{eqnarray}
    k_2 (T) = \sqrt{\frac{2 m}{T} - 1}\,.
\end{eqnarray}
Anticipating the final form of the line element, we identified an integration
constant with (twice) the mass $m$. The other momentum, $k_1$, is
determined by 
\begin{equation}
    k_1 = - \frac{m e_2}{2 T^2 \left( 2 m - T\right)} k_2= -\frac{m
      \sqrt{2m-T}}{2T^{5/2}} e_2
\end{equation}
using the Hamiltonian constraint.
The final equation of motion then implies
\begin{equation}
    \dot{e}_2 = - \mu N \sqrt{1 - s \lambda^2 k_2^2} \left( \frac{e_2 k_2}{\sqrt{e_1}} + 2 \sqrt{e_1} k_1\right) = \frac{1}{2} e_2 \left(\frac{1}{T} - \frac{1}{2 m - T}\right)
\end{equation}
which is solved by
\begin{equation}
    e_2 (T) = \alpha^{-1} \sqrt{T \left(  2 m - T\right)}
\end{equation}
with an integration constant $\alpha$. Note that the free functions $\mu$ and
$\lambda$ canceled out in all the differential equations we had to solve. The
solutions are therefore valid for any $\mu$ and $\lambda$ depending on $T$
through $e_1$.

We now have complete solutions for all phase-space functions and can compute
the lapse function
\begin{equation}
    N = - \frac{1}{\mu \sqrt{\left(2m/T-1\right) \left(1 - s\lambda^2 \left(2 m /T - 1\right)\right)}}
\end{equation}
as well as the emergent radial metric component
\begin{equation}
    q_{X X}^{\rm em} = \frac{e_2^2}{\mu^2 e_1} = \frac{1}{\alpha^2 \mu^2} \left(\frac{2 m}{T}-1\right)\,.
\end{equation}
The space-time metric equals
\begin{equation}
    {\rm d} s_{\rm em}^2
    = - \frac{{\rm d} T^2 / \mu^2 }{(2 m / T-1) (1 +s\lambda^2- 2 m s\lambda^2 / T)}
    + \left( \frac{2m}{T}-1\right) \frac{{\rm d} X^2}{\alpha^2 \mu^2}
    + T^2 {\rm d} \Omega^2
    \label{eq:Homogeneous - Schwarzschild}
\end{equation}
where, without loss of generality, we can absorb the constant $\alpha$ in the
radial coordinate $X$, but we  can keep it to cancel $\mu$ in the case where it is constant.

The range of the time coordinate $T$ is determined by the condition that $N$
is real. In the emergent space-time metric, $N^2$ is always positive and
cannot be split into a sign factor and a lapse function squared as the radial
metric component. To recall, the radial metric component is determined by the structure
function $\gamma$ which may be positive or negative (or zero). According
to the structure of hypersurface deformations, this function determines the
signature of space-time as well as the spatial metric. The lapse function does
not appear in structure functions and therefore can only be used in the
classical form, such that $N^2$ is positive and multiplies ${\rm
  sgn}(\gamma){\rm d}t^2$. 

Using this condition, $T$ has the maximal value $T_{\rm max} = 2m$ at the
horizon as a boundary of the interior region. Its minimum value is
$T^-_{\rm min} = 0$ for $s = -1$, in which case $1-s\lambda^2(2m/T-1)$ in the
lapse function remains positive for all $T$ such that $2m/T>1$. For $s=1$,
there is a positive lower bound on $T$ determined by
\begin{equation}
    T^+_{\rm min} = \frac{2 m \lambda^2}{1+\lambda^2}\,.
\end{equation}
If $\lambda$ is not constant, this is an implicit equation for
$T_{\rm min}^+$. The coordinate chart constructed here then ends, and at least
in the case of constant $\lambda$ it can be extended to an expanding interior
solution as shown in \cite{SphSymmMatter,SphSymmMatter2}.  There is no
signature-change hypersurface in these models because the structure function
$\gamma$ remains positive in the allowed ranges of $T$.
It is now easy to confirm that the line elements (\ref{eq:Homogeneous -
  Schwarzschild}) and (\ref{Schwarzschild}) are indeed related by a simple
flip of space and time coordinates, $T = x$ and $X = t$, using the Lorentzian
solution with $\epsilon=1$ in the latter case.

\subsection{Painlev\'e--Gullstrand line element}
\label{sec:PG gauge}

An interesting gauge choice that allows transitions through the horizon in
classical general relativity is given by the Painlev\'e--Gullstrand
solution. We will first derive a suitable form in emergent modified gravity by
applying coordinate transformations from the exterior and interior solutions
already found, and then confirm that the resulting metric components also
follow uniquely from the constrained system.

\subsubsection{Coordinate transformation from the static Schwarzschild gauge}

In order to derive a suitable coordinate transformation, we first observe that
the metric (\ref{SchwarzschildSig}) has the Killing vector
$\xi_{(t)}^\mu \partial_\mu = \partial_t$.  A timelike geodesic with tangent
vector $u^\mu$ then has the conserved quantity
$e = - g_{\mu \nu} \xi_{(t)}^\mu u^\nu = - u_t$, where $g_{\mu\nu}$
refers to the emergent space-time metric. This equation tells us that $u_t$
always remains finite even if we approach a signature-change hypersurface where
some of the components of $g_{\mu\nu}$ may diverge.

Using the normalization condition
$g_{\mu\nu}({\rm d}x^{\mu}/{\rm d}\tau) ({\rm d}x^{\nu}/{\rm
  d}\tau)=-\epsilon$ for the tangent vector
$u^{\mu}={\rm d}x^{\mu}/{\rm d}\tau$ of a geodesic, with the signature factor
$\epsilon$, a geodesic can be described by the 1-form
\begin{eqnarray}
    \epsilon {\rm d} \tau &=& -u_{\mu}{\rm d}x^{\mu}= - u_t {\rm d} t - u_x {\rm d} x\,.
\end{eqnarray}
A generalization of the classical Painlev\'e--Gullstrand gauge can be defined
by requiring that the new time coordinate $t_{\rm PG}$ equals proper time
along infalling radial geodesics, while the spatial coordinate $x$ remains
unchanged compared with the Schwarzschild solution. If we compute the
components $u_t$ and $u_x$ using normalization and conserved quantities, we
can integrate the resulting ${\rm d}\tau$ and obtain $t_{\rm PG}$ as a
function of $t$ and $x$. Keeping track of the signature factor $\epsilon$,
this construction can be used in Lorentzian and Euclidean regions.

Normalization
\begin{eqnarray}
    - \epsilon &=&
    u_\mu u_\nu g^{\mu \nu}
\end{eqnarray}
with the inverse metric (\ref{eq:Inverse metric - spherical symmetry}) and the
signature factor $\epsilon$ implies
\begin{eqnarray}
     \epsilon \left(- \frac{u_t^2}{N^2} + 1 \right)
    + 2 \epsilon \frac{M}{N^2} u_t u_x
    + \left(q^{x x} + \epsilon \frac{M^2}{N^2}\right) u_x^2
    = 0
    \label{eq:Frame normalization}
\end{eqnarray}
in general. For zero shift in the original space-time, $M=0$, this equation
simplifies to
\begin{eqnarray}
      q^{x x} (u_x)^2
    = \epsilon \left( \frac{u_t^2}{N^2} - 1 \right)
    \label{eq:Frame normalization}
\end{eqnarray}

If there is signature change at large $x$, it is not meaningful to identify
$u_t$ with the conserved energy $e$ as measured by an asymptotic
observer. Instead, we can evaluate the normalization condition at the point
$x_{\lambda}$ of the signature-change hypersurface where
$q^{xx}=0$. Therefore, $u_t^2=N(x_{\lambda})^2$, or
\begin{eqnarray}
    u_t
    = \sqrt{N^2 + \epsilon N^2 q^{x x} (u_x)^2} \big|_{x_0}
    \ ,
\end{eqnarray}
at some reference coordinate $x_0$.
The different values for $u_t$ parametrize the proper time of the different observers.
If we choose an observer at rest at $x_0<x_\lambda$ (where $q^{xx}\not=0$) and
using (\ref{SchwarzschildSig}) we obtain
\begin{eqnarray}
    u_t = - \frac{\epsilon}{\alpha \mu(x_0)} \sqrt{1-\frac{2m}{x_0}}
    \ .
\end{eqnarray}
The sign choice is such that
\begin{equation}
  u^t(x_0) = \frac{\alpha \mu(x_0) }{ \sqrt{1-2m/x_0}}
\end{equation}
is future-pointing.  One can then take the limit $x_0 \to x_\lambda$ for
timelike geodesics that are formally at rest at the signature-change surface.
Only values in the range $x_0\geq x_\lambda$ imply initial values for timelike geodesics in the
Lorentzian region that reach the signature-change hypersurface. (There are no
timelike geodesics starting at $x_0$ if this value is in the Euclidean region,
but we may make a choice $x_0>x_{\lambda}$ just to specify certain initial
values of a timelike geodesic at $x<x_{\lambda}$ that is not at rest anywhere.)
Equation~(\ref{eq:Frame normalization}) then implies
\begin{eqnarray}
    u_x(x)&=& \pm \sqrt{\epsilon q_{xx}(x)\left(\frac{u_t^2}{N(x)^2}-1\right)}\nonumber\\
          &=& s_2 \frac{1}{1-2m/x} \frac{1}{\sqrt{|\lambda(x)^2-1| \,
              |1-X_{\lambda}/x|}} \sqrt{\epsilon\left(\frac{1-2m/x_0}{\mu(x_0)^2}-
              \frac{1-2m/x}{\mu(x)^2}\right)} 
\end{eqnarray}
with $s_2=\pm 1$. Since the signature factor $\epsilon$ changes at
$x=x_{\lambda}$, the square root is real provided $(1-2m/x)/\mu(x)^2$ is
increasing across $x_{\lambda}$, which is the case for any $\mu(x)$ that
does not increase faster than $\sqrt{1-2m/x}$. 

The
coordinate transformation from $t$ to $t_{\rm PG}(t,x)$ is now determined by
\begin{eqnarray} \label{tPG}
    {\rm d} t_{\rm PG} &=& 
    - u_t {\rm d} t- u_x {\rm d}x\\
    &=& \frac{\epsilon}{\sqrt{\alpha} \mu(x_0)} \sqrt{1-\frac{2m}{x_0}}\:{\rm
        d}t\nonumber\\
  &&- s_2  \frac{1}{1-2m/x} \frac{1}{\sqrt{|\lambda(x)^2-1| \, 
              |1-X_{\lambda}(x)/x|}} \sqrt{\epsilon\left(\frac{1-2m/x_0}{\mu(x_0)^2}-
              \frac{1-2m/x}{\mu(x)^2}\right)} \:{\rm d} x \,.\nonumber
\end{eqnarray}
It is impossible to integrate this equation for generic $\lambda(x)$ and
$\mu(x)$, but it is easy to check that the integrability condition
$\partial^2t_{\rm PG}/\partial t\partial x=\partial^2t_{\rm PG}/\partial
x\partial t$ is satisfied.

Writing
\begin{equation}
  {\rm d}t_{\rm PG}= \epsilon N(x_0){\rm d}t- s_2\sqrt{q_{xx}\epsilon
    \left(\frac{N(x_0)^2}{N(x)^2}-1\right)} {\rm d}x\,,
\end{equation}
we obtain
\begin{eqnarray}
  -\epsilon N(x)^2{\rm d}t^2+\epsilon q_{xx}{\rm d}x^2&=& -\epsilon \frac{N(x)^2}{N(x_0)^2} \left({\rm d}t_{\rm PG}+s_2\sqrt{\epsilon q_{xx} \left(\frac{N(x_0)^2}{N(x)^2}-1\right)} \:{\rm d}x\right)^2+ \epsilon q_{xx}{\rm d}x^2\nonumber\\
  &=& -\epsilon {\rm d}t_{\rm PG}^2-
      \epsilon\left(\frac{N(x)^2}{N(x_0)^2}-1\right) {\rm d}t_{\rm
      PG}^2\\
  &&- 2\epsilon s_2 \frac{N(x)^2}{N(x_0)^2} \sqrt{\epsilon q_{xx}\left(\frac{N(x_0)^2}{N(x)^2}-1\right)} {\rm
     d}t_{\rm PG}{\rm d}x+ \epsilon q_{xx}\frac{N(x)^2}{N(x_0)^2}
     {\rm d}x^2 \nonumber\\
  &=& -\epsilon{\rm d}t_{\rm PG}^2+ \epsilon q_{xx}\frac{N(x)^2}{N(x_0)^2}
      \left({\rm d}x - \epsilon s_2
      \sqrt{\epsilon q^{xx} \left(\frac{N(x_0)^2}{N(x)^2}-1\right)} {\rm
      d}t_{\rm PG}\right)^2\,.\nonumber
\end{eqnarray}
The 
emergent line element therefore equals
\begin{eqnarray}
    {\rm d}s^2_{\rm em}
    &=&
    - \epsilon {\rm d} t_{\rm PG}^2
        + \epsilon
        \frac{\mu(x_0)^2}{\mu(x)^4(\lambda(x)^2-1)(1-2m/x_0)(X_{\lambda}(x)/x-1)}\nonumber\\ 
  &&\times
        \left({\rm d}x+ \epsilon s_2 \mu(x)^2 \sqrt{(\lambda(x)^2-1)
        \left(\frac{X_{\lambda}(x)}{x}-1\right) \left(\frac{1-2m/x_0}{\mu(x_0)^2}-
        \frac{1-2m/x}{\mu(x)^2}\right)} {\rm d}t_{\rm PG}\right)^2\nonumber\\
&&    + x^2{\rm d}\Omega^2
    \label{eq:Schwarzschild in timelike observer}
\end{eqnarray}
in a gauge of Painlev\'e--Gullstrand type. Unlike the classical solution in
this gauge, slices of constant $t_{\rm PG}$ are not flat, owing to a
position-dependent factor of $(X_{\lambda}(x)/x-1)^{-1}$. The metric is
degenerate at the signature-change hypersurface, $x=x_{\lambda}$ defined by
$X_{\lambda}(x_{\lambda})=x_{\lambda}$, but not at the horizon $x=2m$ and can
be used in the interior as well as the exterior of the black hole. It is also
well-defined in the Euclidean region $x>x_{\lambda}$, where the square root
remains real (using $\lambda>1$, which is required for signature change to be
possible).

\subsubsection{Painlev\'e--Gullstrand slicing in the canonical theory}

We have obtained the metric in a slicing analogous to the classical
Painlev\'e--Gullstrand gauge by deriving a coordinate transformation from the
exterior Schwarzschild region. Covariance of the underlying theory requires
that the same metric coefficients can be obtained from the canonical equations
with suitable gauge choices. In particular, while $e_1 = x^2$ can still be
used, staticity in the Schwarzschild gauge should be replaced by the condition
of uniform lapse, $N = 1$. The radial component of the emergent line element
and the shift vector are then determined by the canonical equations of motion
and constraints.

A non-vanishing shift vector makes it possible that $k_1$ and $k_2$ are
non-zero even for time-independent $e_1$. The diffeomorphism constraint
implies that these two phase-space functions are related by
\begin{equation}
  k_1= \frac{e_2}{2x}k_2'\,.
\end{equation}
The Hamiltonian constraint then implies a first-order differential equation
relating $k_2$ and $e_2$:
\begin{equation}
  \frac{e_2}{2x}k_2^2+e_2k_2k_2'+\frac{e_2}{2x}-\frac{3x}{2e_2}+
  \frac{x^2}{e_2^2}e_2'=0\,.
\end{equation}
Using $\dot{e}_1=0$ according to one of the gauge conditions, the equation of
motion for $e_1$ with $N=1$ implies
\begin{equation} \label{e1dot}
  Mx+x\mu k_2 \sqrt{1-s\lambda^2k_2^2}+
  s\lambda^2\mu \frac{x^3}{e_2^2}\frac{k_2}{\sqrt{1-s\lambda^2k_2^2}}=0
\end{equation}
from which we obtain the shift vector $M$ as a function of $e_2$ and $k_2$.

We need one additional condition, supplied by the equation of motion for
$e_2$. We assume that this function is time-independent, as in the classical
Schwarzschild solution, but may differ from the classical expression $x$.
Using $e_1=x^2$, the Hamiltonian constraint
(\ref{eq:Modified constraint}) simplifies to
\begin{eqnarray}
  H[1] &=& \int{\rm d}x \mu\sqrt{1 - s \lambda^2 k_2^2} \Bigg(
s \lambda^2 \frac{x^2}{ e_2^2} \frac{k_2(e_2k_2' -2xk_1)}{1 - s \lambda^2 k_2^2} 
    - \frac{e_2 k_2^2}{2x}
    - 2 x k_1k_2 
  \\
  &&\qquad\qquad\qquad\qquad\qquad
    - \frac{x^2e_2'}{e_2^2}
    + \frac{3x}{2 e_2}
    - \frac{e_2}{2 x}\Bigg) \nonumber
\end{eqnarray}
and, with $\dot{e}_2=0$, implies
\begin{eqnarray}
 0&=& s \left(\frac{\mu \lambda^2x^2 k_2^2 }{e_2\sqrt{1-s\lambda^2k_2^2}}\right)'+
      s\frac{\mu \lambda^2x^2 (k_2'-2xk_1/e_2)
      }{e_2\sqrt{1-s\lambda^2k_2^2}}\nonumber\\
  &&
   +   \frac{\mu e_2k_2}{x}\sqrt{1-s\lambda^2k_2^2}+
  2\mu x k_1 \sqrt{1-s\lambda^2k_2^2}+(Me_2)'\nonumber\\
 &=& s \left(\frac{\mu \lambda^2x^2 k_2^2 }{e_2\sqrt{1-s\lambda^2k_2^2}}\right)'+
  \frac{\mu e_2k_2}{x}\sqrt{1-s\lambda^2k_2^2}+
  \mu e_2k_2' \sqrt{1-s\lambda^2k_2^2}+(Me_2)'\,. \label{e2dot}
\end{eqnarray}
Equations~(\ref{e1dot}) and (\ref{e2dot}) contain the combination
\begin{equation}
  f(x)=\frac{\mu e_2\sqrt{1-s\lambda^2 k_2^2}}{x}
\end{equation}
in several places, which may be used instead of $e_2$. Doing so, we obtain
\begin{equation} \label{e1dot2}
  Me_2+ xk_2f+\frac{s\lambda^2\mu^2xk_2}{f}=0
\end{equation}
and
\begin{equation}
 s \left(\frac{\mu^2\lambda^2x k_2}{f}\right)' + (k_2+xk_2')f+ (Me_2)'=0\,.
\end{equation}
Combining the last two equations, several terms cancel out and we arrive at
$f'=0$, such that $f=1/C_2$ is constant and 
\begin{equation}\label{eq: e2 w con}
  e_2(x)=\frac{x}{C_2\mu(x)\sqrt{1-s\lambda(x)^2k_2(x)^2}}\,.
\end{equation}

In the Hamiltonian constraint, we then have the $k_2$-independent contribution
\begin{eqnarray}
  1-3\frac{x^2}{e_2^2}+2\frac{x^3}{e_2^3}e_2'&=&
                                                 1-C_2^2\mu^2(1-s\lambda^2k_2^2)
-2C_2^2x\mu\sqrt{1-s\lambda^2k_2^2} \left(\mu\sqrt{1-s\lambda^2k_2^2}\right)'
\nonumber\\
                                             &=& 1-C_2^2(x\mu^2(1-s\lambda^2k_2^2))'
\end{eqnarray}
which has to equal $-k_2^2-2xk_2k_2'=-(xk_2^2)'$ for the Hamiltonian
constraint to vanish. Therefore,
\begin{equation}
  (x(k_2^2-C_2^2\mu^2(1-s\lambda^2k_2^2)))'=-1
\end{equation}
or
\begin{equation}
  k_2=\pm\sqrt{\frac{C_2^2\mu^2-1+C_k/x}{1+sC_2^2\mu^2\lambda^2}}
\end{equation}
with an integration constant $C_k$, such that $(C_2^2\mu^2-1+C_k/x)/(1+sC_2^2\mu^2\lambda^2)\geq 0$ in a
given range of $x$. We obtain
\begin{equation}
  1-s\lambda^2k_2^2= 
  \frac{1+s\lambda^2(1-C_k/x)}{1+sC_2^2\mu^2\lambda^2}
\end{equation}
and
\begin{equation}
  e_2=\frac{x}{C_2\mu}
  \sqrt{\frac{1+sC_2^2\mu^2\lambda^2}{1+s\lambda^2(1-C_k/x)}}
\end{equation}
provided $x$ is such that
$(1+s\lambda^2-s\lambda^2C_k/x)/(1+sC_2^2\mu^2\lambda^2)\geq 0$.
The shift vector
\begin{equation}
M=- \mu\left(1+\frac{s\lambda^2\mu^2}{f^2}\right) k_2\sqrt{1-s\lambda^2k_2^2}=
s_2 \mu \sqrt{1+s\lambda^2(1-C_k/x)} \sqrt{C_2^2\mu^2-1+C_k/x}
\end{equation}
then follows from (\ref{e1dot2}), where $s_2 = \pm 1$.

Finally, the emergent radial metric is given by
\begin{equation}
  q_{xx}^{\rm em} =   \frac{1}{\mu^2}
  \left|1+ \frac{x^2}{e_2^2} \frac{s\lambda^2}{1-s\lambda^2k_2^2}
    \right|^{-1}
  \frac{e_2^2}{x^2}
= \frac{1}{C_2^2\mu^4} \frac{1}{|1+s\lambda^2-s\lambda^2C_k/x|}
  \end{equation}
with the signature factor
\begin{equation}
    \epsilon = {\rm sgn} (1+ sC_2^2 \mu^2\lambda^2)
    \ ,
\end{equation}
Comparing the radial metric in the case of $s=-1$ with (\ref{eq:Schwarzschild in timelike
  observer}), we identify the free constants as
\begin{equation}
  C_2=\frac{\sqrt{1-2m/x_0}}{\mu(x_0)}
\end{equation}
and
\begin{equation}
  C_k= 2 m \,.
\end{equation}
We arrive at the emergent line element
\begin{eqnarray}
  {\rm d}s_{\rm em}^2
  &=& -\epsilon{\rm d} t_{\rm PG}^2
  + \frac{\mu(x_0)^2}{\mu(x)^4 |1+s\lambda^2| (1-2m/x_0)} \frac{1}{|1-X_\lambda(x)/x|}
  \\
  &&\times \left({\rm d}x
  + s_2 \mu(x)^2 \sqrt{|1+s\lambda^2| \left(1-\frac{X_\lambda(x)}{x}\right)}
     \sqrt{\frac{1-2m/x_0}{\mu(x_0)^2}-\frac{1-2m/x}{\mu(x)^2}} {\rm d} t_{\rm
     GP}\right)^2 \nonumber\\
  &&+x^2{\rm d}\Omega^2
                 \ .\nonumber
\end{eqnarray}

With this choice of constants, motivated by the previous
Painlev\'e--Gullstrand gauge, the signature factor depends on the constant
$x_0$:
\begin{equation}\label{eq:PG signature factor}
    \epsilon = {\rm sgn} \left( 1-\frac{\mu(x)^2}{\mu(x_0)^2} \lambda(x)^2
      \left(1-\frac{2m}{x_0}\right) \right) 
    \ ,
\end{equation}
where we have taken $s=-1$ because only in this case does signature change occur.
For the same reason, we may assume $\lambda^2>1$ in a region around the
signature-change hypersurface.
The conditions given by reality of $k_2$ and $e_2$ can now be rewritten simply as
\begin{eqnarray}
    \left(\frac{1-2m/x_0}{\mu(x_0)^{2}} - \frac{1-2m/x}{\mu(x)^{2}} \right) \epsilon &\geq& 0
    \ , \\
    (\lambda^2-1) (X_\lambda/x-1) \epsilon& \geq& 0
    \ .
\end{eqnarray}
Lorentzian signature $\epsilon=+1$ then requires that $x<x_0$ and
$x<x_\lambda$, while Euclidean signature $\epsilon=-1$ requires that $x>x_0$
and $x>x_\lambda$, provided that $\mu$ does not increase faster than
$1-2m/x$. The choice $x_0=x_{\lambda}$ allows us to connect Lorentzian and
Euclidean signature at the signature-change hypersurface.

\section{Causal structure  near the signature change hypersurface}
\label{s:Causal}

We have obtained covariant models with timelike signature-change hypersurfaces
for $s=-1$ and $\lambda>1$ in a region around the hypersurface. Different
methods such as limiting procedures and coordinate changes can be applied to
elucidate the causal structure of such space(-time)s.

\subsection{Asymptotic behavior}

The metric (\ref{SchwarzschildSig-Lorentzian region}) contains several
$x$-dependent terms that modify some of the classical large-$x$ behavior of
the Schwarzschild solution. Moreover, signature change at $x=x_{\lambda}$
prevents us from taking the full limit of $x\to\infty$ within the Lorentzian
region. A more interesting range is given by $x$ asymptotically close to
$x_{\lambda}$, in which some of the metric factors are approximately
constant.

\begin{figure}
\begin{center}
    \includegraphics[width=15cm]{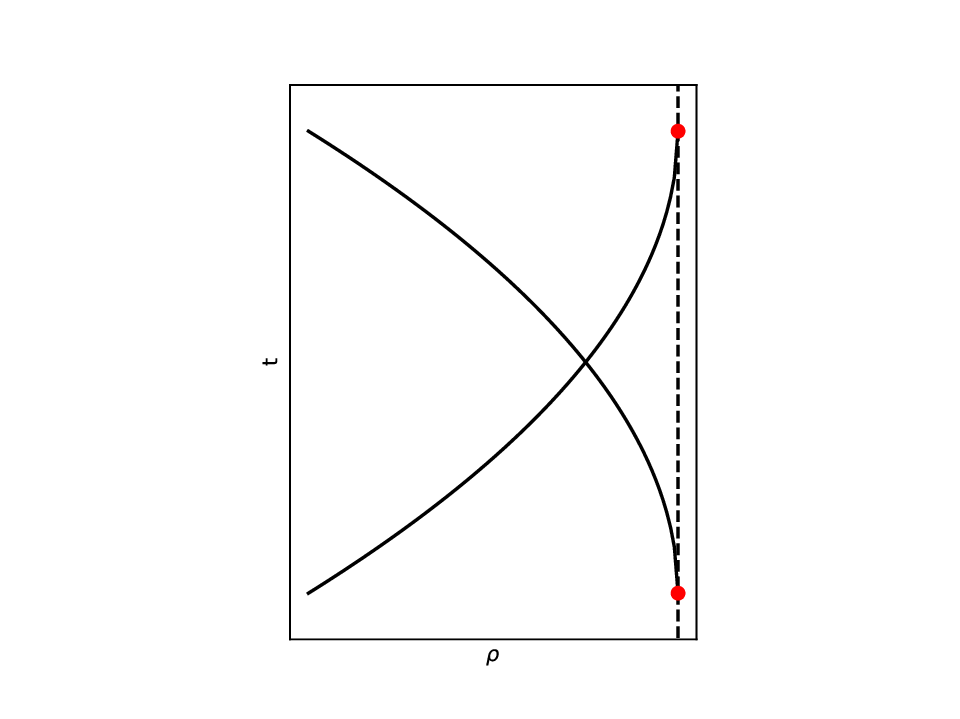}
    \caption{A light cone asymptotically close to the signature-change
      hypersurface (dashed), where light rays end in a tangential direction
      (circles). \label{f:SigChange}}
\end{center}
  \end{figure}

Introducing  the coordinate transformation $x = x_\lambda - \rho$ from $x$ to
a  positive $\rho$, the line element has the asymptotic form
\begin{eqnarray}
    {\rm d}s^2_{\rm em} = - \epsilon \left(1-\frac{2m}{x_\lambda}\right) {\rm d}t^2
    + \epsilon \frac{x_\lambda/(\mu^2(\lambda^2-1))}{(1-2m/x_\lambda)} \frac{{\rm d}\rho^2}{\rho}
    + x_\lambda^2{\rm d}\Omega^2
\end{eqnarray}
with leading-order terms in an expansion by $\rho$. The $\rho$-component of
the metric remains non-constant in these coordinate. However, the line element
becomes manifestly Minkowskian (up to constant scalings of the coordinates)
by defining a new coordinate
\begin{eqnarray}
    X = 2 \sqrt{\rho}
\end{eqnarray}
such that
\begin{eqnarray}
    {\rm d}s^2_{\rm em} = -\left(1-\frac{2m}{x_\lambda}\right) {\rm d}t^2+
  \frac{x_\lambda/(\mu^2(\lambda^2-1))}{(1-2m/x_\lambda)} {\rm d} X^2
  + x_\lambda^2{\rm
    d}\Omega^2
\end{eqnarray}
asymptotically close to the signature-change hypersurface on the Lorentzian
side, $\rho>0$ and $\epsilon=1$.
The same procedure can be used on the other side, $\rho<0$ and
$\epsilon=-1$, instead defining the coordinate $X_{\rm E} = 2 \sqrt{-\rho}$ of
4-dimensional Euclidean space:
\begin{eqnarray}
    {\rm d}s^2_{\rm em} = \left(1-\frac{2m}{x_\lambda}\right) {\rm d}t^2+
  \frac{x_\lambda/(\mu^2(\lambda^2-1))}{(1-2m/x_\lambda)} {\rm d} X_{\rm E}^2
  + x_\lambda^2{\rm
    d}\Omega^2\,.
\end{eqnarray}
These asymptotic geometries can be used to infer the light-cone structure
shown in Fig.~\ref{f:SigChange}.

\subsection{Timelike worldlines}

The local Minkowski form asymptotically close to the signature-change
hypersurface suggests that this hypersurfaces can be reached in finite proper
time from the Lorantzian side. This expectation can be confirmed explicitly by
a derivation of timelike geodesics, and in a similar manner for lightlike
geodesics in the next subsection.

Using the results of Section~\ref{sec:PG gauge}, we can use co-velocity
of a radially infalling object at rest at $x_0$, with components
\begin{eqnarray}
    v_t &=& - \sqrt{1-\frac{2m}{x_0}}
    \ , \nonumber\\
    v_x &=& \frac{1}{\mu\sqrt{\lambda^2-1}} \frac{2 m}{x} \left(1-\frac{2m}{x}\right)^{-1} \left(\frac{X_\lambda}{x}-1\right)^{-1/2} \sqrt{1- \frac{x}{x_0}}
    \ .
\end{eqnarray}
The radial component $v_x$ diverges at the signature change surface unless $x_0=x_{\lambda}$, but
$v^x=q^{xx}v_x$ is finite, and so is ${\rm d} x / {\rm d} t=v^x/v^t$ using
$v^t = - v_t / N^2$:
\begin{eqnarray}
    \frac{{\rm d} x}{{\rm d} t}
    &=& \mu \sqrt{\lambda^2-1} \left(1-\frac{2m}{x}\right) \sqrt{\frac{2
        m}{x}} \sqrt{\left(\frac{X_\lambda}{x} - 1\right) \left(1
        -\frac{x}{x_0} \right)} 
    \ .
    \label{eq:Velocity of timelike geodesic}
\end{eqnarray}
This component  vanishes at $x=x_\lambda$, in agrement with the tangential
approach shown in Fig.~\ref{f:SigChange}.

The proper-time distance along a geodesic starting at some point
$x_i < x_\lambda$ and going up to $x_\lambda$, parameterized by coordinates as
functions of $x$, is given by
\begin{eqnarray}
    \tau_\lambda 
    &=& \int^{x_\lambda}_{x_{i}} \sqrt{- g_{\mu \nu} \frac{\partial x^\mu}{\partial x} \frac{\partial x^\nu}{\partial x}}\: {\rm d} x
    = \int^{x_\lambda}_{x_{i}} \sqrt{- g_{t t} \left(\frac{\partial t}{\partial x}\right)^2
    - 2 g_{t x} \frac{\partial t}{\partial x}
    - q_{xx}}\: {\rm d} x
    \nonumber\\
    &=& \int^{x_\lambda}_{x_{i}} \frac{1}{|v^x|} \sqrt{ N^2 - q_{xx} \left(M
    + v^x\right)^2}\: {\rm d} x
    \ .
\end{eqnarray}

Using the Schwarzschild metric (\ref{SchwarzschildSig-Lorentzian region}),
this equation becomes
\begin{eqnarray}
    \tau_\lambda
    &=&  \int^{x_\lambda}_{x_{i}} \frac{1}{\mu \sqrt{\lambda^2-1}}\sqrt{\frac{x}{2 m}} \left(1-\frac{2m}{x}\right)^{-1/2} \left(\frac{X_\lambda}{x} - 1\right)^{-1/2} \left(1 -\frac{x}{x_0} \right)^{-1/2}
    \nonumber\\
    &&\qquad\qquad
    \times \sqrt{\left( 1 - \left(1-\frac{x}{x_0}\right) \frac{2 m}{x} \right)}\: {\rm d} x
    \ ,
\end{eqnarray}
Consider now the coordinate expansions $x_i = x_\lambda - \rho_i$ and
$x = x_\lambda - \rho$ with positive $\rho_i \ll x_\lambda$ and $\rho$, and
$x_0 > x_\lambda$. To leading order, to which $\mu$ and $\lambda$ may be
treated as constants, we have
\begin{eqnarray}
    \tau_\lambda
    &=& \frac{1}{\mu \sqrt{\lambda^2-1}} \sqrt{\frac{x_\lambda}{2 m}} \left(1-\frac{2m}{x_\lambda}\right)^{-1/2} \left(1 -\frac{x_\lambda}{x_0} \right)^{-1/2} \sqrt{1 - \left(1-\frac{x_\lambda}{x_0}\right) \frac{2 m}{x_\lambda}}
    \nonumber\\
    &&\qquad\qquad
    \times \int_{0}^{\rho_{i}} \sqrt{\frac{x_\lambda}{\rho}} {\rm d} \rho
    \nonumber\\
    &=& \frac{2 x_\lambda}{\mu \sqrt{\lambda^2-1}} \sqrt{\frac{\rho_i}{2 m}} \left(1-\frac{2m}{x_\lambda}\right)^{-1/2} \left(1 -\frac{x_\lambda}{x_0} \right)^{-1/2} \sqrt{1 - \left(1-\frac{x_\lambda}{x_0}\right) \frac{2 m}{x_\lambda}}
    \ ,
\end{eqnarray}
which is finite and and real for $x_0>x_{\lambda}$.
(The value is complex if $x_0<x_{\lambda}$ because geodesics
with such an initial condition do not reach the signature-change
hypersurface.)  In the special case of the most energetic geodesic, formally
given by $x_0 \to \infty$, this result simplifies to
\begin{eqnarray}
    \tau_\lambda
    &=& \frac{2 x_\lambda}{\mu \sqrt{\lambda^2-1}} \sqrt{\frac{\rho_i}{2 m}}
    \ .
\end{eqnarray}
The larger $x_\lambda$ is, the larger this proper time is. 

\subsection{Null worldlines}

For radial null worldlines in (\ref{SchwarzschildSig-Lorentzian region}), we
find a relation between ${\rm d}t$ and ${\rm d}x$ given by
\begin{equation}\label{SchwarzschildSig-Lorentzian region Radial null}
    {\rm d}t = \pm \frac{1}{\sqrt{\mu^2(\lambda^2-1)(X_\lambda/x-1)}}{\frac{{\rm
          d}x}{\left(1 - 2m/x \right)}}\,.
\end{equation}
We can use this result to simplify the co-velocity
\begin{eqnarray}
    {\rm d} \gamma
    &=&
    -v_t {\rm d} t - v_x {\rm d} x
    \nonumber\\
    &=&
    -v_t \left( {\rm d} t + \left(\sqrt{N^2 q^{x x}} + M\right)^{-1} {\rm d} x
        \right)\,, \label{dgamma}
\end{eqnarray}
with a constant $v_t<0$ (such that $v^t>0$) and choosing the negative sign of
$v_x$ in the second term for infalling light rays.

For a timelike worldline, we compute
\begin{equation} \label{dgammatau}
  {\rm d}\gamma = -v_{\nu}{\rm d}x^{\nu}=-g_{\mu\nu} \frac{{\rm d}x^{\mu}}{{\rm
      d}\tau} {\rm d}x^{\nu}= -{\rm d}\tau \left(g_{\mu\nu} \frac{{\rm d}x^{\mu}}{{\rm
        d}\tau} \frac{{\rm d}x^{\nu}}{{\rm d}\tau}\right)= {\rm d}\tau
\end{equation}
and therefore $\gamma$ along the worldline is proper time up to a constant
shift.  The expression ${\rm d}\gamma$ in (\ref{dgamma}), which uses the
geodesic condition through constant $v_t$, can be locally integrated to a
space-time function $\gamma(t,x)$ because it is closed since $v_x$ depends
only on $x$, and therefore locally exact.  The result $\gamma(t,x)$ can be
interpreted as function that determines a foliation of a region of space-time
into curves ${\rm d}\gamma=0$ transversal to timelike geodesics, with a family
of normal directions that integrate to timelike geodesics.

For null worldlines, the analog of the calculation (\ref{dgammatau}) merely
shows that ${\rm d}\gamma=0$ and therefore $\gamma$ is constant along any null
worldline, without providing a relationship with the affine
parameter. Therefore, we do not obtain the affine parameter along null
geodesics as an analog of proper time. Nevertheless, we may use the expression
${\rm d}\gamma$ in order to foliate space-time into null rays, given by
constant $\gamma$ which then plays the role of a null coordinate.  This
foliation allows us to estimate distances to the signature-change hypersurface
as follows: We use one family for null worldlines, given by the infalling
case, in order to introduce $\gamma$ as a null coordinate constant along
infalling null worldlines. (The same null coordinate will be used in order to
transform to Eddington--Finkelstein form in the next subsection.) This
coordinate then provides a certain distance measure along a single outgoing
worldline that approaches the signature-change hypersurface, which corresponds
to the distance one may use to draw a conformal diagram. If this
null-coordinate distance is finite, an observer can send only a finite number
of infalling light rays at regular intervals before reaching the
signature-change hypersurface.

Along a geodesic:
\begin{eqnarray}
    {\rm d} \gamma
    &=&
    -v_t \left( \pm \frac{1}{\sqrt{(\lambda^2-1)(X_\lambda/x-1)}}\frac{1}{\left(1
        - 2m/x \right)} + \left( \sqrt{N^2 q^{x x}} + M\right)^{-1}
        \right) {\rm d}x\,.
    \label{eq:Null coordinates - Kruskal-Szekeres - integrable}
\end{eqnarray}
As before, $N = \mu^{-1}\sqrt{1 - 2m/x}$ and $M = 0$ while
\begin{equation}
    q_{xx} = \frac{1}{(\mu^2(\lambda^2-1))(1-2m/x)(X_{\lambda}/x-1)}
\end{equation}
and thus
\begin{equation}
    q^{xx} = \mu^2(\lambda^2-1)(1-2m/x)(X_{\lambda}/x-1)
\end{equation}
Therefore,
\begin{eqnarray}
    {\rm d} \gamma
    &=&
    \frac{-v_t}{\left(1 - 2m/x \right)\sqrt{(\lambda^2-1)(X_\lambda/x-1)}}
        \Bigg(\pm 1
    + 1 \Bigg){\rm d} x
\end{eqnarray}
vanishes along infalling null wordlines, as expected, and gives us a non-zero
null distance
\begin{eqnarray}
    {\rm d} \gamma
    &=&
    \frac{-2v_t}{\left(1 - 2m/x \right)\sqrt{(\lambda^2-1)(X_\lambda/x-1)}}
       {\rm d} x
    \label{eq:Null coordinates - Kruskal-Szekeres - integrable 4}
\end{eqnarray}
when integrated along outgoing worldlines.

Asymptotically close to the signature-change hypersurface,
$x = x_\lambda - \rho$ with  $0\leq\rho << x_\lambda$, the leading-order expression
\begin{eqnarray}
    \int{\rm d} \gamma
    &=&
    \int\frac{-2v_t}{\left(1 - 2m/x_\lambda \right)\sqrt{(\lambda^2-1)\rho/x_\lambda}}{\rm d} \rho
    \label{eq:Null coordinates - Kruskal-Szekeres - integrable 5}
\end{eqnarray}
can be reduced to
\begin{eqnarray}
    \gamma
    &=&
    \frac{-2v_t}{\left(1 - 2m/x_\lambda \right)\sqrt{(\lambda^2-1)/x_\lambda}} \int_0^{\rho_i} \frac{{\rm d}\rho}{\sqrt{\rho}}
    \label{eq:Null coordinates - Kruskal-Szekeres - integrable 6}
\end{eqnarray}
because  $\lambda$ is approximately constant.
The integral
\begin{eqnarray}
    \gamma
    &=&
    \frac{-4v_t\sqrt{\rho_i}}{\left(1 - 2m/x_\lambda \right)\sqrt{(\lambda^2-1)/x_\lambda}}
    \label{eq:Null coordinates - Kruskal-Szekeres - integrable 6}
\end{eqnarray}
is finite.

\subsection{Null coordinates}

Using a null coordinate $v$, the emergent metric in Eddington--Finkelstein
form is given by
\begin{equation}
    {\rm d} s^2
    =
    - \left(1-\frac{2m}{x}\right) {\rm d} u^2
    + \frac{2}{|\mu| \sqrt{\lambda^2-1}} \frac{1}{\sqrt{X_{\lambda}/x-1}} {\rm d} u {\rm d} x
    \,.
    \label{eq:EF-Sig-Lorentzian region}
  \end{equation}
(For a generic 2-dimensional line element, the Eddington--Finkelstein form is
given by
\begin{equation}
    {\rm d} s^2
    =
    - \frac{N^2 - q_{x x} (N^x)^2}{(v_t)^2} {\rm d} u^2
    + 2 q_{x x} \frac{\sqrt{N^2 q_{x x}}}{v_t} {\rm d} u {\rm d} x
    \label{eq:Eddington-Finkelstein metric form}
\end{equation}
with a null coordinate $u$.)

For constant $\mu$ and $\lambda$, the (outgoing) null coordinate is related to
the original coordinates by a direct integration of ${\rm d}u={\rm d}\gamma$
using equations from the preceding subsection:
\begin{eqnarray}
    u
    &=&
    t
    - \frac{s}{|\mu| \sqrt{|\lambda^2-1|}} \left( \sqrt{x} \sqrt{x_{\lambda}-x}
    + (4 m + x_\lambda) \arctan \left(\sqrt{x_{\lambda}/x-1}\right)
    \right.
    \nonumber\\
    &&\left.
    \qquad
    + \frac{4 m \sqrt{2m}}{\sqrt{x_\lambda-2m}} {\rm arctanh} \left( \sqrt{\frac{2m}{x} \frac{x_{\lambda}-x}{x_\lambda-2m}}\right) \right)
    + c
\end{eqnarray}
with an integration constant $c$, where we have absorbed the constant $v_t$
into the null coordinate.  The coordinate becomes imaginary if we try to
extend it to $x>x_\lambda$, where null worldines no longer exist.  The
modified Eddington-Finkelstein metric (\ref{eq:EF-Sig-Lorentzian region})
still has a coordinate singularity at the signature change surface.

The Eddington--Finkelstein form (\ref{eq:Eddington-Finkelstein metric form})
can directly be transformed to double-null or Kruskal--Szekeres type variables
by introducing
\begin{equation}
  {\rm d}v= \frac{{\rm d}u}{(v_t)^2}+
    - \frac{2N q_{x x}^{3/2}}{N^2 - q_{x x} (N^x)^2}  \frac{{\rm d} x}{v_t}\,.
\end{equation}
In the present case, we obtain
\begin{equation}
    {\rm d} s^2
    =
    - \left(1-\frac{2 m}{x}\right) {\rm d} u {\rm d} v
    \label{eq:KS-Sig-Lorentzian region}
\end{equation}
without modifications from the classical solution. However, the null
coordinates have modified relationships with the Schwarzschild-type
coordinates $x$ and $t$: We have
\begin{equation}
    u = t - x_*
    \quad , \quad
    v = t + x_*
\end{equation}
with
\begin{eqnarray}
    x_*
    &=&
    c + \frac{1}{|\mu| \sqrt{|\lambda^2-1|}} \left( \sqrt{x} \sqrt{x_{\lambda}-x}
    + (4 m + x_\lambda) \arctan \left(\sqrt{x_{\lambda}/x-1}\right)
    \right.
    \nonumber\\
    &&\left.
    \qquad
    + \frac{4 m \sqrt{2m}}{\sqrt{x_\lambda-2m}} {\rm arctanh} \left( \sqrt{\frac{2m}{x} \frac{x_{\lambda}-x}{x_\lambda-2m}}\right) \right)
    \label{eq:x* null coord}
\end{eqnarray}
for constant $\mu$ and $\lambda$.  The null coordinates become imaginary for
$x>x_\lambda$ and hence end at the signature-change hypersurface, even though
(\ref{eq:KS-Sig-Lorentzian region}) does not reveal a coordinate singularity
at this place.

The metric (\ref{SchwarzschildSig}) describing the Euclidean region does not
allow null coordinates in the usual sense.  But if we make the Wick-like
rotation $t \to i \bar{t}$, the metric becomes Lorentzian once again and is
identical to that of the Lorentzian region (\ref{SchwarzschildSig-Lorentzian
  region}) up to the change in time coordinate and retaining a positive radial
component.  Therefore, in these complex coordinates, one can perform the same
procedure as used above in order to obtain null coordinates and a metric of the
Kruskal--Szekeres form.

The result is almost identical, the metric now given by
\begin{equation}
    {\rm d} s^2
    =
    - \left(1-\frac{2 m}{x}\right) {\rm d} \bar{u} {\rm d} \bar{v}
    \label{eq:KS-Sig-Euclidean region}
\end{equation}
where the Schwarzschild coordinates are related to the null ones by
\begin{equation}
    \bar{u} = \bar{t} - \bar{x}_*
    \ , \qquad
    \bar{v} = \bar{t} + \bar{x}_*
\end{equation}
with
\begin{eqnarray}
    \bar{x}_*
    &=&
    c + \frac{1}{|\mu| \sqrt{|\lambda^2-1|}} \left( \sqrt{x} \sqrt{x_{\lambda}-x}
    + (4 m + x_\lambda) \arctan\left( \sqrt{1-x_{\lambda}/x}\right)
    \right.
    \nonumber\\
    &&\left.
    \qquad
    + \frac{4 m \sqrt{2m}}{\sqrt{x_\lambda-2m}} {\rm arctanh} \left( \sqrt{\frac{2m}{x} \frac{x-x_{\lambda}}{x_\lambda-2m}}\right) \right)
    \ .\label{eq:xb* null coord}
\end{eqnarray}
It is therefore possible to draw a single Penrose diagram with both regions
joined at the signature-change hypersurface.

\section{Conclusions}

We have obtained explicit analytical solutions for a large class of
spherically symmetric black-hole models of emergent modified gravity with two
generic modification functions. Focussing on a new type of signature change on
timelike hypersurfaces at low curvature, we have analyzed the causal structure
and confirmed that a Euclidean wall around the universe may be consistent with
astronomical and cosmological observations provided the modification function
$\lambda$ is small in a large range of the radial coordinate $x$ but
eventually crosses the threshold $\lambda=1$ at large distances.

\section*{Acknowledgements}

This work was supported in part by NSF grant PHY-2206591.

%\bibliographystyle{../preprint.bst}
%\bibliography{../Bib/QuantGra.bib}

\end{document}